\documentclass{ckm}                 

\usepackage{txfonts}            
\usepackage{mathptmx}          

\newcommand{\be}{\begin{equation}}
\newcommand{\ee}{\end{equation}}
\newcommand{\bea}{\begin{eqnarray}}
\newcommand{\eea}{\end{eqnarray}}

\newcommand{\gtaeq}{\raisebox{-.6ex}{$\stackrel{\textstyle{>}}{\sim}$}}
\newcommand{\MeV}{\hbox{MeV}}
\newcommand{\GeV}{\hbox{GeV}}
\newcommand{\nf}{n_{\rm{f}}}

\def\msbar{{\rm \overline{MS\kern-0.05em}\kern0.05em}}

\confname{Workshop on the CKM Unitarity Triangle, IPPP Durham, April
  2003}

\title{WG\,V Conveners' Report: Charm Inputs for CKM Physics}

\author{ H. Wittig\addressmark{a}, A.A. Petrov\addressmark{b}, and D.G.
Cassel\addressmark{c}}

\address[a]{DESY, Notkestra{\ss}e 85, D-22603 Hamburg, Germany}
\address[b]{Department of Physics and Astronomy, Wayne State
University, Detroit, MI 48201, USA}
\address[c]{Laboratory for Elementary-Particle Physics, Cornell
University, Ithaca, NY 14853, USA}

\begin{document}

\begin{abstract}
We introduce the contributions to the proceedings of Working Group
V. The main topics were: present and future experiments dedicated to
charm physics, the interplay between high-precision measurements and
better control over theoretical uncertainties, and searches for
signals of new physics in the charm sector. The comparison of Lattice QCD
calculations with precise measurements of charm semi-leptonic and
leptonic decays can have a major impact on the determination of CKM matrix
elements from measurements of $B$ meson decays.

\end{abstract}

\maketitle

\section{Introduction}

Working Group V, dedicated to charm inputs for CKM parameters, was
newly formed for the second workshop in this series. Its formation
reflects the fact that determinations of CKM parameters, as well as
unitarity checks, are increasingly dominated by theoretical
uncertainties in hadronic matrix elements, which cannot be computed in
perturbation theory \cite{ckm02yb}. Although lattice simulations of QCD are
based on first principles, systematic errors in current simulations -- most
notably those due to neglecting the effects of dynamical quarks (quark loops)
-- are still rather large.

The growing importance of charm physics for the study of CKM
parameters is underlined by the decisions to commission future
high-luminosity charm facilities, such as CESR-c and BEPCII.  The availability
of high-precision data in the charm sector will enable the validation of
recent theoretical progress in lattice QCD, aiming at a substantial
reduction of systematic errors \cite{Daviesetal,HasJan,Luscher}.
Direct measurements of the leptonic decay constants $f_D$ and
$f_{D_s}$, as well as semi-leptonic form factors, will challenge the
lattice community to compute these quantities with much greater
precision than currently possible \cite{cleocyb}. Decay constants, form
factors and other quantities can also be computed using QCD sum rules. As for
lattice simulations, the comparison of results obtained in the charm
sector with experiment serves to validate the method.

In this sense, the charm sector serves as a testbed for several
theoretical methods which are needed to fully exploit the data samples
taken at the $B$-factories. In addition to the more supportive r\^ole
of charm physics for the $b$-quark sector, there is also the potential
to discover new physics by studying processes involving charm quarks.

In this report we first discuss several general issues in both theory
and experiment, before detailing recent progress made for a number of
quantities.

\section{Charm Physics: Theory and Experiment}

\subsection{High-Luminosity Charm Facilities}

As noted in the review by D.\,Cassel \cite{dgc_ckm03}, experimental
activity in the charm sector will be greatly boosted by the
CLEO-c/CESR-c programme at Cornell, as well as the proposed BESIII/BEPCII
programme in Beijing. CLEO-c will study $e^+\,e^-$ collisions at
$\sqrt{s}=3-5$ GeV, and it is expected that 1.5 million
$D_s\,\bar{D}_s$ pairs, 30 million $D\,\bar{D}$ pairs, and 1\,billion
$\psi$ decays will be observed. These performance targets imply that
many processes can be studied with unprecedented precision in the
charm threshold region.

The physics programme of CLEO-c will focus on measuring absolute charm
branching fractions, semi-leptonic form factors and the direct
measurement of the leptonic decay constants $f_D$ and $f_{D_s}$. The
expected high precision for these quantities presents a challenge to
lattice QCD to determine them with equal accuracy. In addition, CLEO-c
will look for new physics. In particular, the observation of (large)
CP-violation in charm decays is a clear and unambiguous signature for
new physics. Other signs are related to $D\,\bar{D}$ mixing parameters
and rare charm decays \cite{aap_ckm03,sf_ckm03}. Accurate
determinations of $D\,\bar{D}$ mixing parameters will also provide
additional information which helps to pin down the CKM angle $\gamma$.

\subsection{Lattice QCD}

In the past decade, simulations of QCD on a space-time lattice have
contributed enormously to studies of CKM parameters, by providing
non-perturbative, model-independent information on decay constants,
form factors and $B$-parameters. As suggested by A.\,Buras and emphasised
by P.\,Mackenzie in his plenary talk \cite{pbm_ckm03}, the most
promising strategy to exploit lattice results is to concentrate on
``gold-plated'' quantities, which are easiest for both theory and
experiment. Thus, one-hadron processes involving stable particles,
such as $B\to\pi\ell\nu$ are favoured over $B\to\rho\ell\nu$. In
particular, multi-particle final states with all their complications
due to final-state interactions are (still) very difficult to treat in
simulations.

A lot of experience has been gained in quantifying systematic errors
in lattice simulations. However, most results have so far been
obtained only in the ``quenched approximation'', where quark loops are
neglected in the simulations. Nevertheless, various techniques
developed over the years make it possible to control effects due to
the discretisation and the renormalisation of local operators at the
level of 5\% or better. This implies that systematic errors in lattice
results are dominated by quenching effects, as well as uncertainties arising
from extrapolations to the physical $u$ and $d$ quark masses, from the
region around the strange quark mass. At present the latter issue is
hotly debated \cite{plenary_lat02}, as it strongly affects quantities
like $f_{D_s}/f_D$ and $\xi=f_{B_s}\sqrt{B_{B_s}}/f_B\sqrt{B_B}$,
which directly enter fits of the CKM parameters (see also the
contribution by D. Becirevi\'c to WG\,II \cite{db_ckm03}).

\subsection{QCD Sum Rules}

QCD sum rules \cite{SVZ_79} have been applied over many years to
compute non-perturbative hadronic effects in weak decay amplitudes 
\cite{ColKhod_03}. On the theoretical level, this approach is
complementary to that of lattice QCD. Furthermore, QCD sum rule
predictions for the charm sector can be compared with experimental
data, thereby allowing tests of the stability of sum rules for the
corresponding $b$-decays.

As pointed out in the contribution by A. Khodjamirian \cite{ak_ckm03},
the transition from the $D$- to the $B$-sector is realised in the sum
rule approach, by replacing
\be
  m_c\to m_b,\quad m_D\to m_B,\ldots
\ee
This simple replacement, however, does not mean that the sum rules in
the two quark sectors are equally reliable. In the case of leptonic
decays, the sum rule in the $b$-sector actually turns out to be more
stable. Still, direct comparisons of sum rule predictions with
experimental data and lattice results in the charm sector help to
check the method as such.

\section{Charm lifetimes}

Accurate knowledge of lifetimes of charmed particles is a crucial
ingredient in the conversion of measured relative branching fractions
to partial decay rates, which are obtained by theory. In addition,
precise theoretical predictions of charm lifetimes are important for
the understanding of issues like power-suppressed corrections in
heavy-quark expansions and quark-hadron duality. D. Pedrini, for the
FOCUS collaboration, reviewed their recent measurements of charm
hadron lifetimes \cite{dp_ckm03}, with typical accuracies that exceed
those of the presently quoted PDG averages \cite{PDG02}. Results by
FOCUS have now clearly established the hierarchy of lifetimes in the
mesonic and baryonic sectors:
\bea
& &\tau(D^0)<\tau(D_s^+)<\tau(D^+) \nonumber\\
& &\tau(\Omega_c^0)<\tau(\Xi_c^0)<\tau(\Lambda_c^+)<\tau(\Xi_c^+).
\eea
We note that an accurate measurement of the $D^0$ lifetime is
important for the determination of the lifetime difference in the
$D^0\,\bar{D}^0$ system, and consequently for new physics searches.

\section{Semi-leptonic $D$ decays \label{sec_SL}}

Semi-leptonic decays like $D\to K\ell\nu_\ell$ and
$D\to\pi\ell\nu_\ell$ serve to determine the CKM matrix elements
$|V_{cs}|$ and $|V_{cd}|$. At the workshop, new results for the decays
$D^0\to K^-\ell^+\nu_\ell$ and $D^+\to{\bar{K^0}}^*\ell^+\nu_\ell\to
K^-\pi^+\ell^+\nu_\ell$ from FOCUS were reported by 
D. Pedrini \cite{dp_ckm03}. In the
course of their analysis of the $D\to K^-\pi^+\ell^+\nu_\ell$ decay, a
big forward-backward asymmetry was detected, which can be modelled by
including an additional $S$-wave (see Ref.~\cite{dp_ckm03} for more
details). FOCUS reported results for the relative branching ratio of
$D^+\to K^-\pi^+\ell^+\nu_\ell$ and $D^+\to K^-\pi^+\pi^+$ decays,
including the $S$-wave interference:
\be
  \frac{\Gamma(D^+\to K^-\pi^+\ell^+\nu_\ell)}
       {\Gamma(D^+\to K^-\pi^+\pi^+)} = 0.602\pm0.010\pm0.021.
\ee
As noted in \cite{dp_ckm03}, this number is $1.6\,\sigma$ lower than
CLEO's result, and $2.1\,\sigma$ higher than the E691 measurements.

FOCUS also reported results for ratios of semi-leptonic form factors:
\bea
& & R_V\equiv V(0)/A_1(0)=1.504\pm0.057\pm0.039 \nonumber\\
& & R_2\equiv A_2(0)/A_1(0)=0.875\pm0.049\pm0.064.
\eea
Lattice results for $R_V,\,R_2$ \cite{UKQCD_94,SPQR_lat02} are in
agreement with these numbers, although the error on $R_2$ is
$\gtaeq\;20\%$. 

Semi-leptonic $D\to K\ell\nu_\ell$ and $D\to\pi\ell\nu_\ell$ have been
studied theoretically, using both QCD sum rules and lattice
calculations. Here, the differential decay rates are related to the
CKM matrix elements $|V_{cs}|$ and $|V_{cd}|$ via the form factors
$f_{DK}^+(q^2)$ and $f_{D\pi}^+(q^2)$, respectively. Calculations of
these form factors using light-cone sum rules (LCSR) were reviewed by
A. Khodjamirian \cite{ak_ckm03}. The most recent results compare
favourably with lattice calculations:
\bea
&   f_{D\pi}^+(0) =\;\left\{\begin{array}{ll}
     0.65\pm0.11 & \makebox[3.5em][l]{LCSR} \cite{AKetal_00} \\
     0.57\pm0.06\,(^{+0.01}_{-0.00}) &
                        \makebox[3.5em][l]{Lattice} \cite{SPQR_00}
                        \end{array}\right. \\
&   f_{DK}^+(0) =\left\{\begin{array}{ll}
     0.78\pm0.11 & \makebox[3.5em][l]{LCSR} \cite{AKetal_00} \\
     0.66\pm0.04\,(^{+0.01}_{-0.00}) &
                        \makebox[3.5em][l]{Lattice} \cite{SPQR_00}
                        \end{array}\right.
\eea
Moreover, the resulting integrated decay widths are consistent with
experiment \cite{PDG02}. It should be noted, though, that the LCSR for
$f_{DK}$ depends quite strongly on the value of the strange quark
mass. This underlines the importance of accurate values of the light
quark masses for studies of this kind. Further details are presented
in Section \ref{sec_mcharm}.

CLEO-c will be able to measure semi-leptonic branching ratios with
relative errors of less than about 1\%. This is due to efficient
particle identification and background subtraction. A striking example
for the efficiency is the clean separation of the (Cabibbo suppressed)
decay $D^0\to\pi^-e^+\nu_e$ from the allowed $D^0\to K^-e^+\nu_e$,
whose branching fraction is an order of magnitude larger
\cite{dgc_ckm03}. In order to exploit this level of experimental
precision, the form factors must also be known with an accuracy at the
1\% level. If lattice simulations are able to meet this challenge, the
resulting accuracy on $|V_{cd}|$ and $|V_{cs}|$ will be 2\% or better.

The total decay width of the $D^*$, recently measured by CLEO
\cite{CLEO_width}, allows a determination of the strong coupling
of $D$-mesons to $P$-wave pions, $g_{D^*D\pi}$, which can be compared
with results from lattice simulations and QCD sum rules:
\be
g_{D^*D\pi}=\left\{\begin{array}{ll}
  17.9\pm0.3\pm1.9 & \makebox[3.5em][l]{CLEO}\;\cite{CLEO_width} \\
  18.8\pm2.3\,{}^{+1.1}_{-2.0} &
                           \makebox[3.5em][l]{Lattice}\;\cite{Orsay_02} \\
  10.5\pm3.0                   & \makebox[3.5em][l]{LCSR}\;\cite{LCSR_gddpi}
                   \end{array}\right.
\ee
While there is good agreement between experiment and quenched lattice
calculations, the LCSR result differs significantly. In view of the
fact that LCSRs produce estimates for form factors and decay constants
which agree with lattice predictions, this discrepancy seems rather
puzzling. As noted by A. Khodjamirian \cite{ak_ckm03}, the fairly crude
ansatz of simple quark-hadron duality could be responsible for
this. One possible scenario, pointed out in \cite{Orsay_LCSR02}
proposes the inclusion of a large, negative contribution from radial
excitations to the sum rule, thereby modifying the simple $D^*$-pole
dominance in $D\to\pi\ell\nu_\ell$ decays.

\section{Leptonic $D$ decays}

Measurements of the branching fractions of $D^+\to\ell^+\nu_\ell$ and
$D_s^+\to\ell^+\nu_\ell$ decays can be used to determine
$f_{D^+}|V_{cd}|$ and $f_{D_s}|V_{cs}|$. However, these branching
fractions are rather small: so far only ${\cal
B}(D_s^+\to\ell\nu_\ell)$ has been measured by several collaborations
\cite{PDG02,soldner_hep01}, with ${\cal B}(D^+\to\ell\nu_\ell)$ being
even smaller by an order of magnitude, due to Cabibbo-suppression. For
$B$-mesons, the measurement of the corresponding branching fractions
(at the level expected) is out of reach for the current generation of
experiments.  Leptonic decay constants of heavy-light mesons have, on
the other hand, been computed using lattice QCD and QCD sum rules. In
order to validate these calculations and to assess their reliability
for $B$-meson decays, it is highly desirable to compare with
experimental determinations in the charm sector.

For lattice calculations, the $D_s$ meson is particularly appealing,
since both the charm and the strange quark can be treated directly in
simulations, i.e. no extrapolations are required to make contact with
the physical values of the valence quark masses.

At the workshop, a recent benchmark calculation of $f_{D_s}$ in the
quenched approximation was presented by Rolf \cite{jr_ckm03}. The
quoted value is \footnote{We employ a normalisation in which
$f_\pi=132$ MeV.}.
\be
    f_{D_s}=252\pm9\,\MeV.
\ee
It is worth emphasising that all lattice artefacts have been
eliminated from this result through an extrapolation to the continuum
limit. Therefore, the only uncertainty that remains is due to
quenching. A crude estimate suggests that this could amount to
$-20\,\MeV$, but it is clear that simulations with dynamical quarks
are needed for a reliable quantitative estimate of the effect.

The uncertainties in this calculation can be compared with the
accuracy expected from CLEO-c measurements of $D_s\to\mu\nu$ and
$D_s\to\tau\nu$, about $\pm 1.7$\% for each mode -- assuming $f_{D_s}
= 260$ MeV. These estimates include a common systematic uncertainty
that is about $\pm 1$\%, so averaging the two results will result in
an uncertainty of about $\pm 1.5$\%.

Lattice determinations of $f_D$ involve an extrapolation in the light
quark mass from values around $m_s/2$ down to $m_u,\,m_d$. Recently,
it was pointed out \cite{kr_02} that chiral logarithms introduce a
large uncertainty in estimates of ratios like $f_{B_s}/f_B$ and
$f_{D_s}/f_D$. Although double ratios like
\be
   \left(\frac{f_{B_s}}{f_B}\right)\left/
   \left(\frac{f_{D_s}}{f_D}\right)\right.,
\ee
in which the chiral logarithms largely cancel
\cite{Grinstein,slov_02}, may lead to better estimates for, say,
$f_{B_s}/f_B$, it is implicitly assumed that the region of validity of
Chiral Perturbation Theory extends to quark masses of
$\gtaeq\;m_s/2$. 

Recent sum rule determinations for $f_D$ and $f_{D_s}$ are discussed
in A. Khodjamirian's contribution \cite{ak_ckm03}, and can be summarised
as
\bea
& & f_D = 200\pm20\,\MeV  \nonumber\\
& & f_{D_s}/f_D = 1.11 - 1.27.
\eea
The quoted errors are dominated by the uncertainty in the charm quark
mass. Improvements in the sum rule calculations are discussed in
\cite{ak_ckm03}. 

\section{Charm Quark Mass\label{sec_mcharm}}

Despite not being one of the input parameters that are used directly
in fits to the CKM parameters, the mass of the charm quark is
nevertheless required in several theoretical approaches that
determine, for instance, $f_D$ or semi-leptonic form
factors. Non-perturbative methods must be applied to determine $m_c$
from experimentally accessible quantities such as $m_D$ or the
$J/\psi$ leptonic width.

The status of quark mass determinations using lattice QCD and QCD sum
rules was reviewed at the workshop by R. Gupta
\cite{rg_ckm03}. Recently, several results for $m_c$ obtained from
simulations of quenched QCD have appeared:
\be
   m_c^\msbar(m_c)=\left\{\begin{array}{ll}
      1.301(34)\,\GeV   & \quad\cite{alpha_mc} \\
      1.26(4)(12)\,\GeV & \quad\cite{spqr_mc} \\
      1.33(8)\,\GeV     & \quad\cite{FNAL_mc}
                          \end{array}\right.\,.
\ee
Despite different systematics, these results are in good agreement. In
Refs.~\cite{alpha_mc} and \cite{spqr_mc} the bare charm mass was
related non-perturbatively to the $\msbar$-scheme. Furthermore, the
continuum limit was taken in \cite{alpha_mc}. As in the case of $f_D$
one concludes that the errors within the quenched approximation are
under good control, while a reliable quantitative estimate of the quenching
error must await further studies. In addition to the absolute mass
values, the ratio $m_c/m_s$ was quoted in Ref.~\cite{alpha_mc}:
\be
   m_c/m_s=12.0\pm0.5.
\ee
At the workshop, A. Khodjamirian reviewed QCD sum rule determinations
of $m_c$ (see Table~1 of \cite{ak_ckm03}). Typical results for
$m_c^\msbar(m_c)$ vary between 1.2 and 1.37 GeV, while the quoted
uncertainties for individual calculations range from 20 to
100 MeV. Thus one observes reasonable agreement between QCD sum rules
and lattice calculations.

We have already noted in Section~\ref{sec_SL} that QCD sum rule
determinations of the form factor $f_{DK}^+$ depends strongly on the
strange quark mass. In his review at the workshop Gupta compared
lattice estimates for $m_s$ with recent QCD sum rule
calculations. Lattice results for $m_s$ in the quenched approximation
can be summarised as (see also \cite{hw_lat02})
\be
   m_s^\msbar(2\,\GeV)= 95 - 115\,\MeV,\quad\nf=0,
\ee
where the quoted range largely reflects the systematic effect arising
from choosing different quantities to set the lattice
scale. Simulations with $\nf=2,\,3$ flavours of dynamical quarks yield
$m_s^\msbar(2\,\GeV)= 70 - 90\,\MeV$. Thus, one observes a marked
decrease in $m_s$ when dynamical quark effects are taken into
account. On the other hand, the most recent QCD sum rule estimates
employing the pseudoscalar or scalar sum rules yield
\be
   m_s^\msbar(2\,\GeV)= 100 - 115\,\MeV \;\pm15\%.
\ee
So while the more recent sum rule calculations have resulted in lower
values than had previously been obtained with this method, they are
still not easily reconciled with lattice results for $\nf=2,\,3$. It
should be noted though, that uncertainties are still quite large, and
that dynamical lattice simulations need to mature, before the
situation can be re-assessed.

\section{New Physics}

The Standard Model (SM) is a very constrained system, which implements
a remarkably simple and economic description of all CP-violating
processes in the flavour sector by a single CP-violating parameter, the
phase of the CKM matrix. This fact relates all CP-violating
observables in bottom, charm and strange systems and provides an
excellent opportunity for searches of physics beyond the Standard
Model. Yet, even with the huge amounts of data currently available,
straightforward tests of the Standard Model such as CKM unitarity (for
example, via measurements of the areas of charm unitarity triangles)
could be quite complicated. This is in part due to the smallness of
CP-violating contributions to charm transition amplitudes, which makes
the charm unitarity triangles ``squashed''.

On the other hand, large statistics available in charm physics experiment 
makes it possible to probe small effects that might be generated by the 
presence of new physics particles and interactions. As was discussed by 
A.~Petrov~\cite{aap_ckm03}, a program of searches for New Physics 
in charm is complementary to the corresponding programs in bottom or strange 
systems. This is in part due to the fact that loop-dominated processes such as
$D^0-{\overline D}^0$ mixing or flavour-changing neutral current (FCNC) decays 
are sensitive to the dynamics of ultra-heavy {\it down-type particles}. Also, 
in many dynamical models of New Physics the effects in $s$, $c$, 
and $b$ systems are correlated. 

Even with the FCNC transitions observed in the nearest future, care should 
be taken in the interpretation of the observed transitions, which ultimately 
stems from the fact that the charm quark mass is not far above 
$\Lambda\sim 1$~GeV, the scale of non-perturbative hadronic physics.
A good example is provided by the charm-anticharm mixing studies.The 
current experimental upper bounds on $D^0-{\overline D}^0$ mixing parameters 
$x=\Delta M/\Gamma$ and 
$y=\Delta \Gamma/2\Gamma$ (with $\Delta M$ and $\Delta \Gamma$ being the
mass and lifetime differences of mass eigenstates of $D^0$) are on the order 
of a few times $10^{-2}$~\cite{Nelson}, and are expected to improve in the 
coming years.
One would need high confidence that the Standard Model predictions for
$x$ and $y$ lie well below currently available experimental limits, if 
any future discovery of $D^0-{\overline D}^0$ mixing is to be regarded as
a discovery of New Physics. This is hard, as in the Standard Model mixing
parameters are generated only at the second order in $SU(3)_F$ symmetry 
breaking~\cite{SU3},
\begin{equation}
x,y\sim sin^2\theta_C\times\left[SU(3)~\mbox{breaking}\right]^2.
\end{equation}
Another possible manifestation of new physics interactions in the charm
system is associated with the observation of (large) CP-violation. This 
is due to the fact that all quarks that build up the hadronic states in weak 
decays of charm mesons belong to the first two generations. Since $2\times2$ 
Cabibbo quark mixing matrix is real, no CP-violation is possible in the
dominant tree-level diagrams that describe the decay amplitudes. 
In the Standard Model CP-violating amplitudes can be introduced by including 
penguin or box operators induced by virtual $b$-quarks. However, their 
contributions are strongly suppressed by the small combination of 
CKM matrix elements $V_{cb}V^*_{ub}$. It is thus widely believed that the 
observation of (large) CP violation in charm decays or mixing would be an 
unambiguous sign for new physics. 

Finally, rare decays of $D$-mesons also probe the effects of FCNC. At
the workshop, S.~Fajfer discussed the dominant mechanisms in radiative
$D$-decays and their potential to detect new physics
\cite{sf_ckm03}. Transitions like $c\to{u}\gamma$ are strongly
suppressed in the SM, with branching ratios estimated at
$3\times10^{-8}$ \cite{c2ugam}. However, in the MSSM gluino exchange
can enhance the branching ratio by two orders of magnitude
\cite{spdw}. A similar enhancement occurs in transitions like
$c\to{u}\ell^+\ell^-$, whose branching ratios in the SM are at the
level of $10^{-10}$. The amplitudes of rare $D$-decays are typically
dominated by long-distance effects. The most promising decay channels
to search for new physics are $D^0\to\rho\gamma$ and
$D^0\to\omega\gamma$, which could be detected at CLEO-c. The two
branching ratios can be combined so as to mostly cancel the
long-distance effects. Another possibility is the $B_c\to B_u^*\gamma$
decay, for which long-distance contributions are expected to be much
smaller. It should be added, though, that the observation of rare
decays will be extremely hard experimentally.

\section*{Acknowledgements}
H.W. would like to thank all contributors to WG\,V for discussions
which were essential in preparing this report. A.A.P acknowledges
invaluable discussions with friends and collaborators, as well as
financial support of the National Science Foundation and the US
Department of Energy. D.G.C. acknowledges the contributions of his
CLEO and CESR colleagues and the support of the National Science
Foundation. All of the authors give special thanks to the organizers
and their staff who worked so hard to make this workshop pleasant and
successful.

\def\etal{{\it et al.}}

\end{document}